\documentstyle[psfig]{mn}

\def\simg{\mathrel{\rlap{\raise 0.511ex \hbox{$>$}}{\lower 0.511ex \hbox{$\sim$}}}}
\def\siml{\mathrel{\rlap{\raise 0.511ex \hbox{$<$}}{\lower 0.511ex \hbox{$\sim$}}}}
\def\etal{et al$.$ } \def\eg{e$.$g$.$ } \def\ie{i$.$e$.$ }

\def\reference{\bibitem}

\begin{document}

\title[ GRB 990123 ]
  {A unified picture for the $\gamma$-ray and prompt optical emissions of GRB 990123}

\author[Panaitescu \& Kumar ]{A. Panaitescu$^1$ \& P. Kumar$^2$ \\
$^1$ Space Science and Applications, Los Alamos National Laboratory, Los Alamos, NM 87545, USA \\
$^2$ Department of Astronomy, University of Texas, Austin, TX 78712 }

\maketitle

\begin{abstract}
\begin{small}
  The prompt optical emission of GRB 990123 was uncorrelated to the $\gamma$-ray light-curve and 
exhibited temporal properties similar to those of the steeply-decaying, early X-ray emission observed
by Swift at the end of many bursts. These facts suggest that the optical counterpart of GRB 990123
was the large-angle emission released during (the second pulse of) the burst. If the optical and 
$\gamma$-ray emissions of GRB 990123 have, indeed, the same origin then their properties require 
that 
(i) the optical counterpart was synchrotron emission and $\gamma$-rays arose from inverse-Compton 
      scatterings (the "synchrotron self-Compton model"), 
(ii) the peak-energy of the optical-synchrotron component was at $\sim 20$ eV, and
(iii) the burst emission was produced by a relativistic outflow moving at Lorentz factor 
      $\simg 450$ and at a radius $\simg 10^{15}$ cm, which is comparable to the outflow 
      deceleration radius.  
Because the spectrum of GRB 990123 was optically thin above 2 keV, the magnetic field behind the 
shock must have decayed on a length-scale of $\siml 1\%$ of the thickness of the shocked gas, 
which corresponds to $10^6-10^7$ plasma skin-depths. 
Consistency of the optical counterpart decay rate and its spectral slope (or that of the burst, 
if they represent different spectral components) with the expectations for the large-angle burst
emission represents the most direct test of the unifying picture proposed here for GRB 990123.
\end{small}
\end{abstract}


\section{Introduction}
\label{intro}

 The Swift satellite has evidenced the existence (in a majority of bursts) of a fast-decaying 
phase after the end of $\gamma$-ray emission (\eg O'Brien \etal 2006), during which the 0.3--10
keV flux falls-off as $F_x \propto t^{-(1.5-4)}$. The emission of the optical counterpart of 
GRB 990123, measured by ROTSE (Akerlof \etal 1999), has a similarly steep decay, $F_o \propto 
t^{-(1.5-2.5)}$ at about the same time (50--400 s after trigger) as the fast-decaying phase of 
Swift X-ray afterglows. This similarity suggests that the optical counterpart of GRB 990123 and 
the fast decay phase of Swift X-ray afterglows originate from the same shock of the GRB relativistic 
outflow. In Swift bursts, the transition from the prompt emission to the fast X-ray decay is 
continuous, which indicates that the $\gamma$-ray and X-ray emissions also have a common origin. 
This leads to the conjecture that the optical and $\gamma$-ray emissions of GRB 990123 arise
from the same shock.

 As shown in figure 2 of Galama \etal (1999), the prompt optical emission of GRB 990123 is well 
above the extrapolation of the burst continuum to lower energies. Then, if the optical and burst
emission originate from the same part of the outflow, they must represent different spectral 
components, \ie optical must be synchrotron emission and $\gamma$-ray must be inverse-Compton 
scatterings. This is the "synchrotron self-Compton model" which has been used by Panaitescu \& 
M\'esz\'aros (2000) and Stern \& Poutanen (2004) to explain the hard low-energy spectra observed 
for some BATSE bursts (Preece \etal 1998). Kumar \etal (2006) have shown that the temporal and 
spectral properties of GRBs 050126 and 050219A favour this model for the $\gamma$-ray emission.

 As discussed by Nousek \etal (2006) and Zhang \etal (2006), the fast-decay phase of Swift 
afterglows can be identified with the "large-angle emission" released during the burst, \ie the 
emission from the fluid moving at an angle $\theta$ larger than the inverse of the outflow's 
Lorentz factor $\Gamma$, with $\theta$ measured relative to the outflow center -- observer axis. 
Any radiating GRB outflow whose opening is larger than $\Gamma^{-1}$ yields a large-angle emission, 
irrespective of the type of shock (internal, reverse-external, or forward-external) and radiative 
process. As shown by Kumar \& Panaitescu (2000), relativistic effects lead to a simple relation 
between the spectral slope and decay index of the fast-decaying X-ray afterglow, which is generally 
found to be consistent with the observations. 

\begin{figure}
\centerline{\psfig{figure=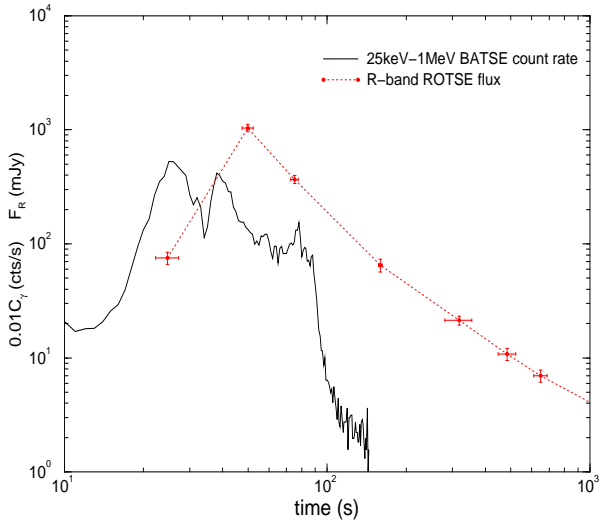,height=7cm,width=8cm}}
\caption{ The BATSE 25--1000 keV count-rate for GRB 990123, with 1 s time resolution, 
          and the ROTSE $R$-band light-curve of its optical counterpart. }
\label{lc}
\end{figure}

 If, as argued above, the prompt optical emission of GRB 990123 and the fast-decay phase of
Swift X-ray afterglows have the same origin, then the former could also be identified with the
large-angle emission produced during the burst. This conjecture can explain why the optical
emission of GRB 990123 appears uncorrelated with that at $\gamma$-rays.
As shown in Figure \ref{lc}, the optical emission of GRB 990123 is weaker during the first pulse, 
exhibits a maximum during the tail of the second pulse (which peaks at 38 s), and then decays 
monotonically throughout the third GRB pulse and after the burst end.
The decoupling of the optical and $\gamma$-ray emissions of GRB 990123 can be explained if the 
optical counterpart is identified with the large-angle emission released during the second GRB 
pulse (when the optical counterpart peaks) and if the optical emission of other pulses is weaker 
than that of the second GRB pulse. Given the simple structure of GRB 990123 light-curve and the 
sparse sampling of the optical counterpart, the lack of an optical-$\gamma$-ray temporal correlation 
could also be the result of fluctuations in the optical-to-$\gamma$-ray output ratio from pulse 
to pulse. Thus, the large-angle emission is not a unique explanation for the uncorrelated optical 
and burst emissions of GRB 990123; it just represents a possible reason and a working assumption
for the calculations below.

 Based on the above arguments, in this work we attribute the optical prompt emission of GRB 990123 
to the large-angle synchrotron emission produced during the second GRB pulse and identify the prompt
$\gamma$-ray emission with up-scatterings of the synchrotron photons. The observational constraints
imposed on this scenario are presented in \S\ref{obs} and used in \S\ref{model} to determine the 
outflow parameters which accommodate them. In \S\ref{extra}, we discuss some implications of the
large-angle emission scenario for the optical counterpart and a possible shortcoming of the 
synchrotron self-Compton model, which can be circumvented if the magnetic field decays and does
not fill the entire GRB outflow. 

 We emphasize two aspects of the following treatment of the unification of the $\gamma$-ray and 
prompt optical emissions of GRB 990123. 

 First, we do not assume a certain mechanism for the dissipation of the relativistic outflow energy. 
This mechanism could be $(i)$ internal shocks in an unsteady wind, as proposed by M\'esz\'aros \& 
Rees (1999), or $(ii)$ the external reverse-shock, as proposed by Sari \& Piran (1999) (fig. 1 of 
Panaitescu \& M\'esz\'aros 1998 also shows that a 10--16th magnitude optical emission could arise 
from the reverse-shock), and further investigated by Kobayashi \& Sari (2000), Soderberg \& 
Ramirez-Ruiz (2002), Fan \etal (2002), Panaitescu \& Kumar (2004), Nakar \& Piran (2005), and 
McMahon, Kumar \& Piran (2006). Thus, the scenario proposed here does not represent a new theoretical
framework for the GRB emission. 

 Second, the calculations below address primarily the implications of the proposed unifying scenario
and represent a test of that scenario only to the extent that the resulting physical parameters are 
plausible. Otherwise, the proposed scenario for the optical counterpart of GRB 991023 is motivated 
by $(i)$ the similarity between its temporal properties and those of the X-ray emission following 
Swift bursts, and $(ii)$ the identification of the latter with the large-angle burst emission. 
The only observational test for the proposed scenario is the consistency between the decay of
the optical emission of GRB 990123 and the expectation from the large-angle emission for the
measured low-energy slope of the burst spectrum (\S\ref{obs}).

\section{Optical and Gamma-ray emissions of GRB 990123}
\label{obs}

\begin{figure}
\centerline{\psfig{figure=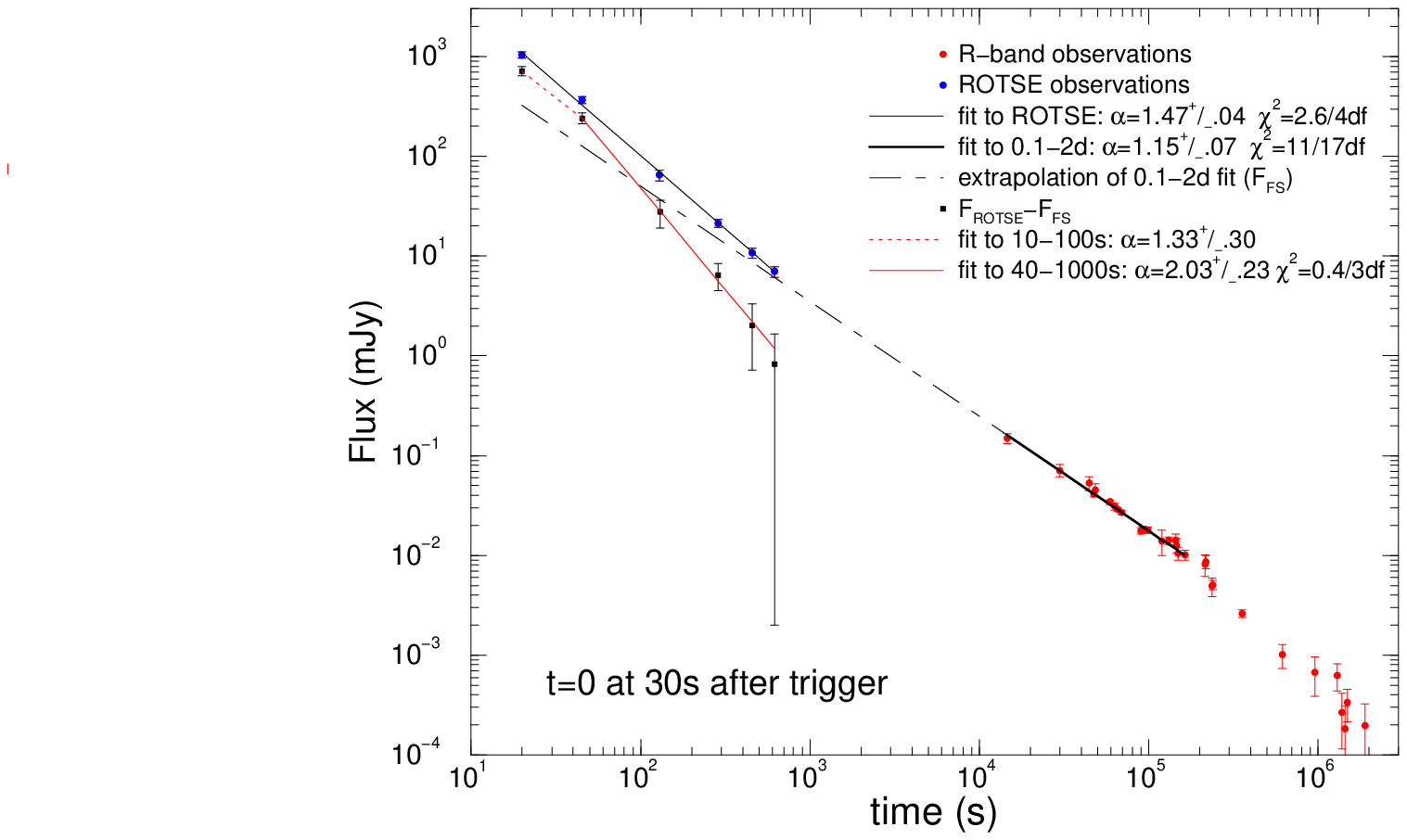,height=7cm,width=9cm}}
\caption{ The 0.1--2 day optical light-curve of GRB afterglow 990123 (red symbols), 
   fit with a power-law decay (thick, solid line), extrapolated to the epoch of ROTSE 
   observations (blue symbols) and subtracted to isolate the early optical emission in 
   excess of that from the forward-shock (black symbols). This excess emission exhibits
   an increasing decay rate after 50 s. Time is measured from 30 s after trigger,
   which is when the second GRB pulse starts (this pulse peaks about 8 s after its beginning
   and 12 s before the largest optical flux measured by ROTSE). The excess optical emission 
   can arise in the same mechanism that generated the burst, its continuation after the 
   end of the burst (which is at 60 s) being due to the large-angle prompt emission.  }
\label{t30}
\end{figure}

 The optical measurements of GRB 990123, shown in figure \ref{lc}, are too sparse to pinpoint 
when the flux peaked, but sufficient to show that a substantial fraction of the post-peak optical 
flux arose during the second GRB pulse. As burst emission episodes may be dynamically independent, 
the decay of the ROTSE optical emission should be timed from the onset of the second GRB pulse, 
which occurred at $\sim 30$ s after the GRB trigger, as shown in Figure \ref{t30}. The optical 
counterpart emission decays as a power in time $F(t) \propto t^{-\alpha}$, with index $\alpha 
\simeq 1.5$. 

 The available optical coverage of Swift afterglows indicates that, quite often, the optical 
emission decays as a power-law from the first observations, at only 100--200 s after trigger 
(see figure 1 of Panaitescu \etal 2006). This suggests that the forward-shock contributes to 
the optical emission just at the end of the burst and motivates us to back-extrapolate the 
0.1--2 day optical emission of GRB afterglow 990123 to the epoch of the ROTSE observations and 
subtract it to determine the optical emission at that time which is excess of the forward-shock 
contribution. As shown in Figure \ref{t30}, the ROTSE excess emission has a decay index 
$\alpha = 1.8 \pm 0.1$, but the power-law fit is not that good, having $\chi^2 = 6.4$ for 4 
degrees of freedom. The reason is that the excess emission exhibits a steepening decay, from 
$\alpha = 1.3 \pm 0.3$ for the first two measurements after the second GRB pulse to $\alpha = 
2.0 \pm 0.2$ during the last four measurements (figure \ref{t30}). 

 If the decay of the ROTSE optical flux is indeed the large-angle emission produced during 
the burst and if this emission switches-off sufficiently fast, then the slope $\beta_o$ of 
the optical spectral energy distribution (SED), $F_\nu \propto \nu^{\beta_o}$, must be
\begin{equation}
 \beta_o = 2 - \alpha 
\label{alpha}
\end{equation}
(Kumar \& Panaitescu 2000). Thus, the above possible decay indices $\alpha$ of the ROTSE
optical emission imply that $0 < \beta_o < 0.7$. The consistency of $\beta_o$ with the burst 
SED slope at low-energy (20--300 keV), which Briggs \etal (1999) report to be $\beta_{LE} 
= 0.4 \pm 0.1$, supports the synchrotron self-Compton interpretation for the optical and 
$\gamma$-ray emissions of GRB 990123 because, in this model, the SED of synchrotron and 
inverse-Compton components must have the same spectral slopes at frequencies above 
self-absorption. Therefore, the SED of the emission of GRB 990123's optical counterpart 
should be
\begin{equation}
  F_\epsilon \propto \left\{ \begin{array}{ll}    
        \nu^{1/3}          & \epsilon < \epsilon_{p,sy}  \\ 
        \nu^{-\beta_{HE}}  & \epsilon > \epsilon_{p,sy}
      \end{array} \right. 
\label{syspek}
\end{equation}
as the only expected spectral slope consistent with $\beta_o$ and $\beta_{LE}$ is 1/3.

 $\beta_o = 1/3$ corresponds to the optical range being above the self-absorption energy, 
$\epsilon_{a,sy}$, but below the peak-energy of the synchrotron spectrum, $\epsilon_{p,sy}$.
$\beta_{LE} = 1/3$ implies that the 20--300 keV range is between $\epsilon_{a,ic} = 
\gamma_p^2 \epsilon_{a,sy}$ and $\epsilon_{p,ic} = \gamma_p^2 \epsilon_{p,sy}$, where 
$\gamma_p$ is the peak Lorentz factor of the electron distribution with energy in the 
shocked fluid and $\epsilon_{p,ic}$ is the peak-energy of the burst spectrum. 

 The low-energy spectrum of GRB 990123, $F_\nu \propto \nu^{-\beta_{LE}}$ peaks at an energy 
which is a fraction $\beta_{LE}/(\beta_{LE}+1)$ of the peak-energy $E_p$ of the $\nu F_\nu$ 
spectrum. According to Briggs \etal (1999), $E_p \simeq 720$ keV for the second GRB pulse, 
therefore 
\begin{equation}
 \epsilon_{p,ic} (t_p) = 210\; {\rm keV} \;.
\label{epic}
\end{equation}

 The burst SED at high-energy (1--10 MeV) is a power-law of slope $\beta_{HE} = -2.1 \pm 0.1$
(Briggs \etal 1999). This shows that the $\gamma_p$-electrons at the peak of the power-law 
electron distribution with energy, 
\begin{equation}
 \frac{dN}{d\gamma}(\gamma > \gamma_p) \propto \gamma^{-p} 
\label{dndg}
\end{equation}
do not cool radiatively because, in the opposite case, the distribution of cooled electrons, 
$dN/d\gamma \propto \gamma^{-2}$, would yield a much harder GRB spectrum ($F_\nu \propto 
\nu^{-1/2}$) above its peak energy. Barring a chance situation where $\gamma_p$ is equal
to the Lorentz factor $\gamma_c$ above which all electrons undergo a significant radiatively
cooling, this shows that $\gamma_c^2 \epsilon_{c,sy} > 10$ MeV, where $\epsilon_{c,sy}$ is 
the synchrotron energy at which the $\gamma_c$-electrons radiate. Then, $\epsilon_{p,ic} = 
\gamma_p^2 \epsilon_{p,sy} = 210$ keV, $\epsilon_{p,sy} \propto \gamma_p^2$, and 
$\epsilon_{c,sy} \propto \gamma_c^2$, lead to $\gamma_c/\gamma_p > (10^4/210)^{1/4}$:
\begin{equation}
 \gamma_c \simg 2.6 \, \gamma_p \;.
\label{gc}
\end{equation}
 Furthermore, the observed high-energy burst spectral slope implies that the electron 
distribution index is $ p = 2\beta_{HE} + 1 \simeq 5.2$ .

 The emission released at a radius $r$ by some fluid patch moving at an angle $\theta$ 
relative to the direction toward the observer arrives at observer at time
\begin{equation}
 ct = \frac{1}{2} r (\theta^2 + \Gamma^{-2})
\label{rt}
\end{equation}
where $\Gamma$ is the bulk Lorentz factor of the GRB-emitting source, and is Doppler-boosted 
by a factor 
\begin{equation}
 {\cal D} = \frac{2\Gamma}{\Gamma^2 \theta^2 + 1} \;.
\label{D}
\end{equation}
As discussed in \S\ref{intro}, in the large-angle emission interpretation for the optical 
counterpart of GRB 990123, the optical emission is released during the second GRB pulse 
(which peaks at time $t_p=8$ s after the beginning of the second GRB pulse). For a source
whose emission switches-off instantaneously, the emission received at $t_p$ comes from the 
fluid moving at angle $\theta = \Gamma^{-1}$ relative to the direction toward the observer. 
From equations (\ref{rt}) and (\ref{D}), that emission arrives at $t_p = r/(c\Gamma^2)$ and
is boosted by ${\cal D} (t_p)= \Gamma$. After $t_p$, emission arrives from $\theta > 
\Gamma^{-1}$, for which $t/t_p = (\Gamma^2\theta^2 + 1)/2$ and 
\begin{equation}
 {\cal D}(t) = {\cal D}(t_p) \frac{t_p}{t}
\end{equation}
on virtue of equations (\ref{rt}) and (\ref{D}). 

 Therefore, the large-angle emission arriving at fixed observer frequency corresponds 
to an ever-increasing comoving-frame frequency. Then, the above conclusion that, for GRB 990123, 
optical is below the peak-energy of the synchrotron spectrum ($\epsilon_{p,sy}$) 
implies that the synchrotron light-curve should steepen at $t_+$ when $\epsilon_{p,sy}$ 
crosses the optical domain:
\begin{equation}
 2\,{\rm eV} = \epsilon'_{p,sy} {\cal D}(t_+) = \epsilon'_{p,sy} {\cal D}(t_p) 
               \frac{t_p}{t_+} = \epsilon_{p,sy} \frac{t_p}{t_+}
\end{equation}
where prime denotes a quantity in the comoving-frame. 
The ROTSE emission with the forward-shock contribution subtracted (Figure \ref{t30}) shows 
such a steepening at about 45 s, the subsequent decay index, $\alpha \simeq 2$, implying an
optical SED slope $\beta_o \simeq 0$, as expected at the peak of synchrotron spectrum. 
For now, we parameterize the time when $t_+$ relative to the epoch of the second GRB pulse 
peak: $t_+ = x\, t_p$. Therefore, the observer-frame synchrotron peak-energy is
\begin{equation}
 \epsilon_{p,sy} (t_p) = 2\,x\; {\rm eV} \quad x > 1 \;.
\label{epsy}
\end{equation}

 The extrapolation of the power-law fit to the ROTSE light-curve to $t_p$, predicts an 
optical flux of $F_R (t_p) = 4$ Jy. The extrapolation of the forward-shock (FS) emission 
to the same time is $F_{FS} (t_p) = 0.93$ Jy, therefore the peak of the excess optical 
emission (which we attribute to the same mechanism as the burst itself) is $F_{sy} (t_p) 
= F_R(t_p)-F_{FS}(t_p) = 3.07$ Jy (the forward-shock emission may have started later 
than the optical peak time, in which case the $F_{sy} (t_p)$ above underestimates the 
true synchrotron flux by 25\%). Taking into account that the slope of the optical SED
is $\beta_o = 1/3$, it follows that the flux at the peak of the synchrotron spectrum is
\begin{equation}
  F_{p,sy} (t_p) = F_{sy} (t_p) \left( \frac{ \epsilon_{p,sy} }{ 2\,{\rm eV} } \right)^{1/3} 
                  = 3.1\, x^{1/3} \,{\rm Jy} \;.
\label{fpsy}
\end{equation}

 The flux $F_{p,ic}$ at the peak-energy $\epsilon_{p,ic}$ keV of the inverse-Compton 
spectrum can be derived from the 100 keV flux of 0.61 mJy reported by Briggs \etal (1999) 
at 17 s after the onset of the second GRB pulse and from the burst flux decrease by about 
40\% from $t_p=8$ s to 17 s: 
\begin{equation}
  F_{p,ic} (t_p) = (210/100)^{\beta_{LE}}\times 1.75 F_{100k}(17\,s) \simeq 1.5\; {\rm mJy} \;.
\label{fpic}
\end{equation}

\begin{figure}
\centerline{\psfig{figure=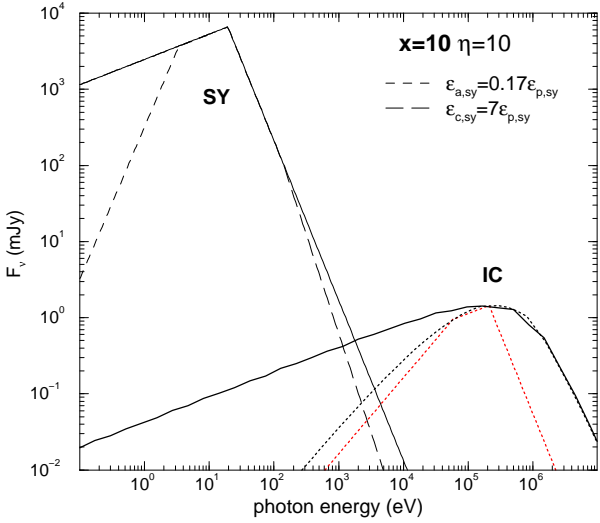,height=7cm,width=8cm}}
\caption{ Synchrotron (optical) and inverse-Compton ($\gamma$-ray) components for GRB 990123. 
  The synchrotron SED is the simplified spectrum given in equation (\ref{syspek}) and has the
   peak frequency and flux given in equations (\ref{epsy}) and (\ref{fpsy}) for $x=10$. 
  The inverse-Compton spectrum is calculated by integrating the scattered emissivity per
   electron over the electron distribution (equation \ref{dndg}). 
  Dotted curves show the effect of self-absorption for $\epsilon_{a,sy} = 0.17 \epsilon_{p,sy}$,
   (equation \ref{easy}). 
  Dashed line shows the effect of electron cooling for $\epsilon_{c,sy} = 7 \epsilon_{p,sy}$ 
   (equation \ref{gc}). 
  For comparison, the dot-dashed curve shows the inverse-Compton spectrum for monoenergetic 
   electrons with $\gamma_p$ given in equation (\ref{gp}).
    }
\label{spek}
\end{figure}

 Figure \ref{spek} shows schematically the synchrotron and inverse-Compton SED, the characteristics 
of the former having been derived above in the large-angle emission interpretation of the optical 
counterpart, while those of the latter come directly from observations.

 Equations (\ref{epic}), (\ref{gc}), (\ref{epsy}), (\ref{fpsy}) and (\ref{fpic}) represent 
the conditions which we use to constrain the physical parameters of the synchrotron 
self-Compton model for the ROTSE optical and BATSE $\gamma$-ray emissions of GRB 990123. 
An upper limit on the parameter $x$ is obtained if the synchrotron power-law spectrum above 
$\epsilon_{p,sy}$ extends up to the $\gamma$-ray range, \ie there are no other spectral breaks 
but $\epsilon_{c,sy}$, by requiring that the synchrotron flux at 2 keV (the lowest $X$-ray 
observational frequency) does not exceed the inverse-Compton flux:
\begin{equation}
 F_{p,sy} \left( \frac{ \epsilon_{i,sy} }{ \epsilon_{c,sy} } \right)^{\beta_{HE}}
          \left( \frac{ \epsilon_{c,sy} }{ 2\; {\rm keV} } \right)^{\beta_{HE}+0.5} <
 F_{p,ic} \left( \frac{ 2\; {\rm keV} }{ \epsilon_{p,ic} } \right)^{\beta_{LE}}
\end{equation}
which leads to $x < 14$ for $\epsilon_{c,sy} = 7\epsilon_{p,sy}$, which is the lowest
value of the cooling energy (see equation \ref{gc}), and to $x < 9$ for $\epsilon_{c,sy} 
> 2$ keV. Therefore, under the assumption that the synchrotron spectrum extends up keV 
energies, the peak-energy of the synchrotron spectrum should satisfy
\begin{equation}
 x \siml 7-14
\label{xgamma}
\end{equation}
and should cross the optical at $t_+ = x\,t_p < 115$ s after the peak of the second GRB pulse 
(145 s after trigger). If ROTSE emission is mostly the large-angle emission of the second 
GRB pulse then the optical light-curve should exhibit a break at $t_+$, which is consistent 
with observations (see Figure \ref{t30}). Based on the constraint above, we normalize
$x$ to 10.

\section{Synchrotron and inverse-Compton emissions}
\label{model}

 Integrating the scattering photon spectrum per electron (\eg equation 2.48 of Blumenthal 
\& Gould 1970) over the synchrotron spectrum (equation \ref{syspek}) and over the electron 
distribution (equation \ref{dndg}), we obtain that 
\begin{equation}
 \epsilon_{p,ic} = 0.82\; \gamma_p^2\, \epsilon_{p,sy} \;.
\label{eicsy}
\end{equation}
Therefore, the typical electron Lorentz factor in the shocked fluid is
\begin{equation}
 \gamma_p=\left( \frac{\epsilon_{p,ic}}{0.82\,\epsilon_{p,sy}} \right)^{1/2}= 110\, x_1^{-1/2} \;,
\label{gp}
\end{equation}
where $x_1=x/10$ (the notation convention $Q_n = Q/10^n$ will be use hereafter).

 The observed peak-energy of the synchrotron spectrum is 
\begin{equation}
 \epsilon_{p,sy} = \frac{4}{3}\, \Gamma\, \frac{\epsilon'_{p,sy}}{z+1}
\end{equation}
where the factor 4/3 accounts for the flux-weighted average frequency (from equation \ref{D},
Doppler boost decreases from ${\cal D} = 2\Gamma$ at $\theta=0$ to ${\cal D} = \Gamma$ at 
$\theta = \Gamma^{-1}$ and goes asymptotically to zero for $\theta \rightarrow \pi$), 
$z$ is the burst redshift and 
\begin{equation}
 \epsilon'_{p,sy} = \frac{3\psi(p)}{4\pi}\, \frac{e h}{m_e c}\; \gamma_p^2\, B
\end{equation}
is the comoving-frame peak-energy of the synchrotron spectrum, $B$ being the magnetic field 
in the shocked fluid and $\psi(p)$ a factor which a weak dependence on the electron index
$p$. By integrating the synchrotron emissivity per electron over the power-law electron 
distribution, Wijers \& Galama (1999) find that $\psi(5.2) = 0.34$, therefore
\begin{equation}
 \epsilon_{p,sy} = 3.2 \times 10^{-9} \, \gamma_p^2 B\, \Gamma\;  {\rm eV} \;.
\end{equation}
Using equation (\ref{gp}) and the observed $\epsilon_{p,sy}$, this leads to
\begin{equation}
  B \Gamma = 4.9 \times 10^5\, x_1^2\;\; {\rm G} \;.
\label{BG}
\end{equation}

 The received flux at the peak of the synchrotron spectrum is
\begin{equation}
 F_{p,sy} = \frac{z+1}{4\pi D_L^2(z)}\, \Gamma\, L'_{p,sy}
\end{equation}
where $D_L(z)$ is the burst luminosity distance, the factor $\Gamma$ accounts for relativistic
beaming of the burst emission (over the region $\theta < \Gamma^{-1}$, the specific flux is 
beamed by a factor $\Gamma^3$, but that regions has an area which is a fraction $\Gamma^{-2}$ 
of the entire emitting surface, assuming spherical symmetry) and 
\begin{equation}
 L'_{p,sy} = \sqrt{3} \phi(p) \frac{e^3}{m_e c^2}\, B N 
\end{equation}
is the comoving-frame luminosity at $\epsilon'_{p,sy}$ and $N$ is the number of radiating
electrons (for a spherically symmetric outflow). The factor $\phi(p)$ calculated by Wijers 
\& Galama (1999) is $\phi(5.2)=0.70$. 

 The observed GRB and optical flux at any time may be the superposition of many ($\eta$) 
emission episodes (sub-pulses resulting from \eg internal collisions in the outflow) in 
which a smaller number ($N_e$) of electrons radiate: $N = \eta\, N_e$. The superposition 
of these sub-pulses leads to intrinsic fluctuations in the burst light-curve, of relative 
amplitude $\eta^{-1/2}$. If the observing time resolution were shorter than the sub-pulse 
duration then the amplitude of the GRB light-curve fluctuations (which includes Poisson noise 
in addition to the source fluctuations) would represent an upper limit for the amplitude of
the intrinsic source fluctuations. In that case, the less than 10\% fluctuations of GRB 990123 
displayed in figure 1 of Briggs \etal (1999) indicate that $\eta \simg 100$. However, the 
sub-pulse duration which we obtain below, of about $\delta t = 20$ ms, is a factor 10 less 
than the typical $t_{res} = 256$ ms resolution of BATSE light-curves. In this case, there 
would be $N_{sp} = \eta (t_{res}/\delta t) \simeq 10\, \eta$ sub-pulses in a GRB pulse.
The resulting amplitude of source fluctuations, $N_{sp}^{-1/2}$, is upper bound by the 
observed $\siml 10\%$ fluctuations of GRB 990123, hence $\eta \simg 10$ suffices to accommodate 
the observed fluctuations. Normalizing $\eta$ to 10, the synchrotron peak flux is
\begin{equation}
  F_{p,sy} = 2.5 \times 10^{-53}\;\eta_1 \,B\,\Gamma\,N_e   \; {\rm mJy} \;.
\label{ffpsy}
\end{equation} 
The observed synchrotron peak flux $F_{p,sy}$ and $B\Gamma$ from equation (\ref{BG}) lead to
\begin{equation}
  N_e = 5.4 \times 10^{50}\; x_1^{-5/3}\,\eta_1^{-1} \;.
\label{Ne}
\end{equation}

 Because the synchrotron and inverse-Compton spectra are similar, the ratio $Y$ of the
inverse-Compton and synchrotron radiating powers is
\begin{equation}
 Y = \frac{\epsilon_{p,ic}\; F_{p,ic}}{\epsilon_{p,sy}\;  F_{p,sy}} = 
     2.4\; x_1^{-4/3} = 0.82\, \gamma_p^2 \frac{F_{p,ic}}{F_{p,sy}} 
\label{Y}
\end{equation}
using equation (\ref{eicsy}).
 At the same time, the Compton parameter is the integral over the electron distribution
of the average increase in the photon energy through scattering:
\begin{equation}
 Y = \int_{\gamma_p}^\infty \frac{4}{3} \gamma^2 \frac{d\tau_e}{d\gamma} d\gamma = 
     \frac{4}{3} \frac{p-1}{p-3}\, \gamma_p^2\, \tau_e = 2.54\, \gamma_p^2\, \tau_e 
\label{YY}
\end{equation}
where $\tau_e$ is the sub-pulse optical thickness to electron scattering and $d\tau_e/d\gamma
\propto dN/d\gamma \propto \gamma^{-p}$.
The above two equations lead to
\begin{equation}
 \tau_e = 0.32\, \frac{F_{p,ic}}{F_{p,sy}} = 7.1\times 10^{-5}\, x_1^{-1/3} 
\end{equation}
using the synchrotron and peak fluxes of GRB 990123.
The electron optical thickness is set by the electron column density: 
\begin{equation}
 \tau_e = \frac{\sigma_T}{4\pi}\, \frac{N_e}{r^2}
\label{taue}
\end{equation}
where $\sigma_T$ is the Thomson cross-section.
The last two equations allow the determination of the radius a which the burst and prompt optical
emissions are released:
\begin{equation}
 r_\gamma = 6.3 \times 10^{14}\; x_1^{-2/3}\, \eta_1^{-1/2} \; {\rm cm} \;.
\label{r}
\end{equation}

 The last observational constraint to be used is equation (\ref{gc}). The radiative cooling
timescale of the $\gamma_p$-electrons is
\begin{equation}
  t_c (\gamma_p) = \frac{z+1}{\Gamma}\, t_c'(\gamma_p) = 
                   6\pi (z+1) \frac{m_e c}{\sigma_T}\, \frac{1}{\gamma_p B^2 \Gamma (Y+1)}
\end{equation}
where $t'_c (\gamma_p)$ is the comoving-frame cooling timescale.
Substituting $B$ from equation (\ref{BG}) and $Y$ from equation (\ref{Y}), leads to
\begin{equation}
 t_c (\gamma_p) = 3.1\times 10^{-5}\, x_1^{-13/6}\, \Gamma \;\; {\rm s} \;.
\label{tc}
\end{equation}
Because $t_c \propto \gamma^{-1}$, the cooling timescale for the $\gamma_c$-electrons,
$t_c(\gamma_c)$, is a factor $\gamma_c/\gamma_p > 2.6$ smaller than the above $t_c(\gamma_p)$.

 The lack of the signature of electron cooling in the spectrum of GRB 990123 means that the 
time $t_\Delta$ that the electrons spend radiating is smaller than $t_c(\gamma_c)$. 
The $t_\Delta$ represents the lifetime of the magnetic field in the shocked fluid. An estimate
of it is provided by the time it takes the shocks to propagate through the shells that generate 
the burst emission. Shell spreading\footnotemark
\footnotetext{ The origin of the burst emission from a shell with this thickness is the only 
  non-trivial assumption made in this section}
is expected to yield a comoving-frame shell-thickness $\Delta' \sim r/\Gamma$, which is crossed 
by a relativistic shock in an observer-frame time 
\begin{equation}
 t_\Delta =  (z+1)\frac{\Delta'}{c\Gamma} = 
             5.7 \times 10^4\, x_1^{-2/3}\, \eta_1^{-1/2}\, \Gamma^{-2} \; {\rm s}
\label{tDelta}
\end{equation}
for a shell at the GRB radius $r_\gamma$ given equation (\ref{r}).
Then, the condition $t_c (\gamma_c) > t_\Delta$ leads to a lower limit on the Lorentz factor 
of the shocked fluid:
\begin{equation}
 \Gamma \ge 1660 \; x_1^{1/2}\, \eta_1^{-1/6} \;.
\label{Gm}
\end{equation}

 Substituting in equation (\ref{BG}), we find a lower limit on the magnetic field
\begin{equation}
 B \le 290\; x_1^{3/2}\, \eta_1^{1/6}\; {\rm G}
\end{equation}
while equation (\ref{tDelta}) yields an upper limit on the sub-pulse duration:
\begin{equation}
 \delta t \le 20\; x_1^{-5/3}\, \eta_1^{-1/6}  \; {\rm ms} \;.
\end{equation}
Note that $r$, $\Gamma$, $B$ and $\delta t$ have a weak dependence on the somewhat uncertain 
number $\eta$ of overlapping sub-pulses within a GRB pulse.  and that the timescale for 
the source intrinsic fluctuations, $\delta t$, is smaller than the temporal resolution 
(256 ms) of GRB 990123 light-curve shown in figure 1 of Briggs \etal (1999), \ie the source 
intrinsic fluctuations are averaged over 10 times the fluctuation timescale. 

 The comoving-frame peak-energy of the inverse-Compton spectrum is  
$\epsilon'_{p,ic} = (z+1) \epsilon_{p,ic}/\Gamma \siml 7\times 10^{-4}\, m_e c^2$, hence
$\gamma_p \epsilon'_{p,ic} \siml 0.07 m_e c^2$. This means that the second inverse-Compton 
scattering occurs around the Klein-Nishina reduction in the scattering cross-section.
Integrating the Compton parameter given in equation (\ref{YY}) over the burst spectrum and
electron distribution, and using the Klein-Nishina scattering cross-section, we obtain that
the Compton parameter for the second scattering is $\tilde{Y} \simg 0.89\, \gamma_p^2 \tau_e
= 0.82\,x_1^{-4/3}$, \ie a factor $\siml 3$ lower than $Y$ for the first scattering. 
Therefore, the synchrotron self-Compton model implies the existence of a high-energy 
component, whose $F_\nu$ spectrum peaks at $\gamma_p^2 \epsilon_{e,ic} \sim 2\, x_1^{-1}$ 
GeV and having a fluence $\Phi_{GeV} = \tilde{Y} \Phi_\gamma \simg 3\times 10^{-4}\, 
{\rm erg\; cm^{-2}}$.

\section{Implications of synchrotron self-Compton model for GRB 990123}
\label{extra}

\subsection{Jet edge}

 If the optical counterpart of GRB 990123 is indeed the large-angle emission produced during 
the second GRB pulse (\ie if there is little contribution to the ROTSE optical emission from 
subsequent GRB pulses) then the optical light-curve should exhibit a sharp break when photons 
emitted from the edge of the jet ($\theta = \theta_{jet}$) arrive at the observer. From equation 
(\ref{rt}), this break should be seen at
\begin{equation}
 t_{edge} = (z+1)\, \frac{r_\gamma\,\theta_{jet}^2}{2\,c} \;, 
\label{tedge}
\end{equation}
which depends on the yet-undetermined jet opening.

 The forward-shock optical emission of the afterglow 990123 exhibits a steepening at $t_{jet} 
\siml 3$ d (Figure \ref{t30}). If attributed to a collimated outflow (Kulkarni \etal 1999), 
then the jet opening is 
\begin{equation}
 \theta_{jet} = \frac{1}{\tilde{\Gamma}(t_{jet})}
\label{tjet}
\end{equation}
where $\tilde{\Gamma}$ is the Lorentz factor of the circumburst medium swept-up by the 
forward-shock, which can be calculated from the isotropic-equivalent kinetic energy $E$ of the 
shock and the density of the circumburst medium: 
\begin{equation}
 E = 4\pi\, r\,A\,m_p c^2 \tilde{\Gamma}^2 \;.
\label{adb}
\end{equation}
The above equation applies for a circumburst medium with a radial proton density distribution 
$n(r) = A r^{-2}$, characteristic for the wind ejected by a massive star as the GRB progenitor, 
and results from that the comoving-frame energy-per-particle in the shocked medium is $\tilde{\Gamma}$. 
For the above the dynamics of the shocked medium, $\tilde{\Gamma} \propto r^{-1/2}$, integration 
of the equation for photon arrival 
\begin{equation}
 t = \frac{3(z+1)}{2c}\, \int^r \frac{dr'}{\tilde{\Gamma}^2(r')}   
\end{equation}  
where the factor 3 relates $\tilde{\Gamma}$ to the arrival time of photons emitted from $\theta = 
\tilde{\Gamma}^{-1}$ (as most of the afterglow emission arises from this location), leads to 
\begin{equation}
 \tilde{\Gamma}(t) = \left( \frac{3\,E}{16\pi\, c\, A\, m_pc^2} \right)^{1/4} 
                    \left( \frac{t}{z+1}\right)^{-1/4} \;.
\end{equation}
Taking the GRB output $E_\gamma = 3\times 10^{54}$ ergs as an estimation of the forward-shock 
kinetic energy and parameterizing the wind medium density to that of a massive star
ejecting $10^{-5}\,{\rm M_\odot\, yr^{-1}}$ at $10^3\,{\rm km\; s^{-1}}$, \ie $A = 3\times
10^{35} A_*\, {\rm cm^{-1}}$, we obtain
\begin{equation}
 \tilde{\Gamma} (t_d) = 25\, E_{54.5}^{1/4} A_*^{-1/4}\, t_d^{-1/4}
\label{GFS}
\end{equation}
where $t_d$ is observer time measured in days.

\begin{figure}
\centerline{\psfig{figure=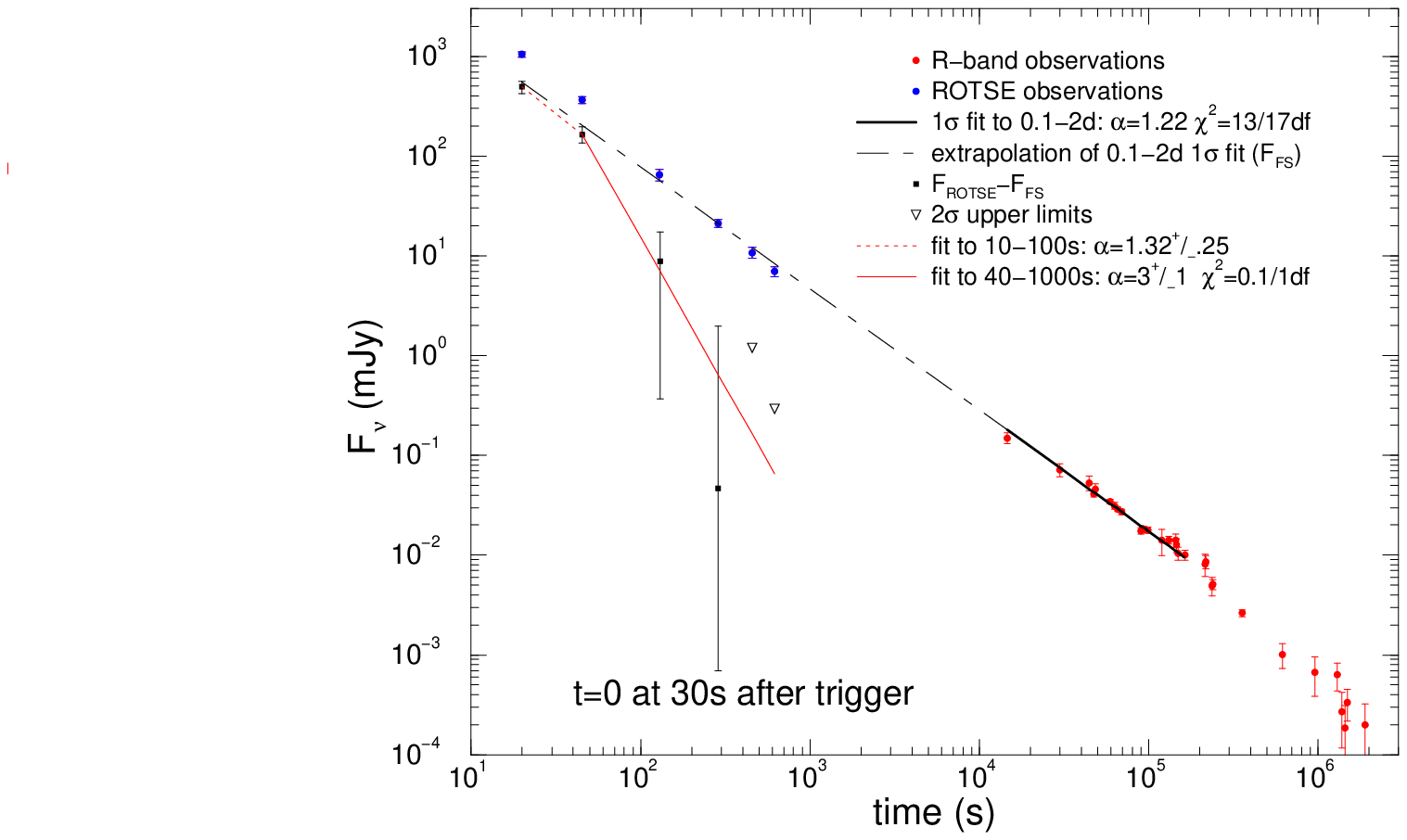,height=7cm,width=8.5cm}}
\caption{ Same as in Figure \ref{t30}, but using a steeper fit, $F_{FS} \propto t^{-1.23}$
   instead of $F_{FS} \propto t^{-1.23}$. The ROTSE light-curve with the forward-shock
   emission subtracted exhibits now a sharp decline at $t \simg 50$ s. }
\label{t30a}
\end{figure}

 From equations (\ref{tjet}) and (\ref{GFS}), the jet opening corresponding to a light-curve
break at $t_{jet} \siml 3$ d is
\begin{equation}
 \theta_{jet} \siml 0.052\, E_{54.5}^{-1/4} A_*^{1/4} \; {\rm rad} \;.
\end{equation}
Then, from equation (\ref{tedge}) and (\ref{r}), the large-angle emission should end at
\begin{equation}
  t_{edge} \siml 74\; x_1^{-2/3} \eta_1^{-1/2} E_{54.5}^{-1/2} A_*^{1/2} \; {\rm s}  \;.
\label{ttedge}
\end{equation}
 Coincidentally, this is about the same as the time $t_+ = xt_p = 80\, x_1$ s when the peak 
energy of the synchrotron spectrum crosses the optical, \ie there are two independent factors
which imply the existence of a steeper decay of the large-angle emission after 80 s from the 
beginning of the second GRB pulse (110 s after trigger).
 This conclusion is at odds with the ROTSE light-curve shown in Figure \ref{t30} but is
consistent with the optical counterpart emission after subtracting a slightly more steeply-decaying
forward-shock contribution, as illustrated in Figure \ref{t30a}. The power-law fit to the 
0.1--2 day optical light-curve shown in Figure \ref{t30a} is statistically acceptable ($\chi^2 = 
13/17$ dof) and has a decay index $\alpha$ larger than that of the best-fit shown in Figure 
\ref{t30} by $1\sigma$.

\subsection{Self-absorption photon energy}

 For the power-law electron distribution given in equation (\ref{dndg}), the self-absorption 
linear coefficient at frequency $\nu$ is
\begin{equation}
  \alpha_\nu  = \frac{p+2}{8\pi}\, \frac{n_e}{(\gamma_p m_e c)^2} \, \nu^{-2}
    \int_{\gamma_p}^\infty d\gamma P_\nu(\gamma) \left( \frac{\gamma}{\gamma_p} \right)^{-(p+1)} 
\label{anu}
\end{equation}
where $n_e$ is the electron density and 
\begin{equation}
 P_\nu (\gamma) = \frac{\sqrt{3}\pi}{4}\, \frac{e^3 B}{m_e c^2}\, F\left(\frac{\nu}{\nu_c}\right)  
   \;\;,\;\;\; \nu_c = \frac{3}{16}\, \frac{e}{m_e c}\, B \gamma^2
\end{equation}
is the synchrotron specific emissivity per electron, $F (x) = x \int_x^\infty K_{5/3} (\xi)\, {\rm d}\xi$ 
being the synchrotron function ($K_{5/3}$ is the modified Bessel function of index 5/3) and $\nu_c$ 
the synchrotron characteristic frequency for electrons of energy $\gamma m_e c^2$.

 At $\nu \ll \nu_c (\gamma_p)$, the synchrotron function is
\begin{equation}
 F(y) \simeq 2.15\, y^{1/3}
\label{Fy1}
\end{equation}
and the integral in equation (\ref{anu}) can be calculated analytically, leading to the
following optical thickness to synchrotron self-absorption, $\tau_a(\nu) = 
(\alpha_\nu \tau_e)/(n_e\sigma_T)$, 
\begin{equation}
  \tau_a(\nu) = 3.3\, \frac{p+2}{p+2/3}\,\frac{e\,\tau_e}{\sigma_T B\gamma_p^5} 
            \left[ \frac{\nu}{\nu_c(\gamma_p)} \right]^{-5/3}   \;.
\label{taua}
\end{equation}
For $p=2.5$ and the parameters $B$, $\tau_e$ and $\gamma_p$ derived in the section 
\S\ref{model}, the synchrotron optical thickness at photon energy $\epsilon$ is
\begin{equation}
  \tau_a(\epsilon) \simg 5.1 \times 10^{-2}\, x_1^{2/3} \eta_1^{-1/6} \;
            \left( \frac{\epsilon}{\epsilon_{p,sy}} \right)^{-5/3}   \;.
\end{equation}
Therefore, the self-absorption energy of the synchrotron spectrum, defined by 
$\tau_a(\epsilon_{a,sy}) = 1$, is 
\begin{equation}
 \epsilon_{a,sy} \simg 0.17\; x_1^{2/5} \eta_1^{-1/10}\, \epsilon_{p,sy} 
                 = 3.3\; x_1^{7/5} \eta_1^{-1/10} \, {\rm eV} \;.
\label{easy}
\end{equation}

 If $\epsilon_{a,sy}$ were above the optical, then $\beta_o = 2$ and the large-angle emission 
would be flat (from equation \ref{alpha}). That the ROTSE optical light-curve decays, implies 
that $\epsilon_{a,sy}$ is below the optical domain. Then equation (\ref{easy}) leads to
\begin{equation}
  x \siml 10\; \eta_1^{1/14} \;.
\end{equation}
which is close to the upper limit obtained by requiring that the synchrotron flux does not 
overshine the inverse-Compton emission at 2 keV (equation \ref{xgamma}). 
The synchrotron self-absorption energy given in equation (\ref{easy}) implies that the 
up-scattered self-absorption energy is
\begin{equation}
 \epsilon_{a,ic} \simg \frac{4}{3} \, \gamma_p^2 \epsilon_{a,sy} =
                   57\, x_1^{2/5} \eta_1^{-1/10} \;{\rm keV}  \;,
\label{eaic}
\end{equation}
where the $(4/3)\, \gamma_p^2$ factor is the average increase of the up-scattered photon energy.

 BeppoSAX observations of GRB 990123 (Corsi \etal 2006) have shown that, at the epoch of the 
first two ROTSE measurements shown in Figure \ref{t30}, the burst $F_\nu \propto \nu^{1/3}$ 
spectrum extends down to 2 keV, \ie a for an other 1.5 dex in energy below the $\epsilon_{a,ic}$ 
obtained in equation (\ref{eaic}).
The up-scattered spectrum below $\epsilon_{a,ic}$ is $F_\nu \propto \nu$ (and not $\propto \nu^2$, 
as for the synchrotron spectrum below $\epsilon_{a,sy}$), thus the synchrotron self-Compton model 
flux at 2 keV flux would be a factor $\sim 30$ below that observed by BeppoSAX. 
A reduction by a factor 2 of that factor is obtained by integrating the up-scattered radiation 
over the synchrotron and electron distribution which, as shown in Figure \ref{spek}. 
We conclude that, for $\epsilon_{a,ic} \simg 57$ keV, the numerically calculated synchrotron 
self-Compton model flux at 2 keV is a factor $\simg 10$ below that without self-absorption
(\ie below the BeppoSAX flux) and that a reduction of $\epsilon_{a,ic}$ by a factor larger
than 10 is required to bring the synchrotron self-Compton model in accord with observations.

\subsection{Decaying magnetic field}

 The above difficulty encountered by the synchrotron self-Compton model and the large-angle
emission interpretation of the ROTSE counterpart, \ie the high up-scattered self-absorption 
energy, is alleviated if the magnetic field does not fill the entire shocked region 
(of radial length $\Delta')$, but decays to a negligible value at some distance $b \Delta'$
($b < 1$) behind the shock which energize the emitting fluid. This implies a reduction of the 
synchrotron flux $F_{p,sy}$ (equation \ref{ffpsy}) by a factor $b$, which must be compensated 
by increasing the number of electrons $N_e$ (equation \ref{Ne}) by a factor $b^{-1}$, as
the product $B\Gamma$ is fixed by the peak-energy of the synchrotron spectrum (equation \ref{BG}). 
 The Compton parameter $Y$ and the minimum electron Lorentz factor $\gamma_p$ remain the same, 
because they are the $\gamma$-ray-to-optical fluences ratio (equation \ref{Y}) and the 
square-root of the $\gamma$-ray-to-optical peak-energies ratio (equation \ref{gp}), respectively,
hence the optical thickness $\tau_e$ of all electrons (within and outside the region filled 
with magnetic field) is unchanged. 

 Consequently, the emission radius $r \propto (N_e/\tau_e)^{1/2}$ (equation \ref{taue}) 
increases by a factor $b^{-1/2}$. This means that the time $t_{edge} \propto r$ when the 
jet edge is seen (equation \ref{tedge}) increases by the same factor $b^{-1/2}$ and that 
large-angle emission can last longer than given in equation (\ref{ttedge}).

 As for self-absorption, a decaying magnetic field means that the column density of the
electrons embedded in the magnetic field (\ie the electrons which absorb the synchrotron
flux) is a fraction $b$ of the total electron column density, hence $\tau_a$ of equation
(\ref{taua}) is multiplied $b$ and the up-scattered self-absorption energy $\epsilon_{a,ic} 
\propto \tau_a ^{3/5}$ by $b^{3/5}$. 
The dependence of $\tau_e$ on the filling factor $b$ is slightly different if the Lorentz 
factor is at the lower limit implied by the condition $\gamma_c \simg 2.6\, \gamma_p$ (which
lead to equation \ref{Gm}) and the magnetic field at the upper limit corresponding to equation 
(\ref{BG}). If the electron radiative cooling is synchrotron dominated ($Y < 1$) then the time 
electrons spend in the magnetic field and cool becomes $b t_\Delta$ (equation \ref{tDelta})
while, if scatterings dominate ($Y > 1$), the electron cooling timescale (equation \ref{tc})
becomes $t_c(\gamma_p)/b$ because the intensity of the synchrotron emission to be up-scattered
is $b$ times lower. Thus, in either case, the condition $\gamma_c \simg 2.6\, \gamma_p$ for 
electron cooling during the burst leads to $t_c(\gamma_p) > 2.6\, b\, t_\Delta$, consequently
the lower limit on the Lorentz factor $\Gamma$ (equation \ref{Gm}) decreases by a factor $b^{-1/6}$
and the upper limit on $B$ resulting from equation (\ref{BG}) increases by a factor $b^{-1/6}$.
It follows that, if the magnetic field strength $B$ is at its upper limit, $\epsilon_{a,ic}$ 
of equation (\ref{eaic}) gets multiplied by a factor $b^{7/10}$, which is close to the $b^{3/5}$ 
factor inferred above for the case when $\Gamma$ above its lower limit.

 Thus, for
\begin{equation}
  b \siml 0.05\, x_1^{-4/7}\, \eta_2^{1/7} 
\label{bmax}
\end{equation}
the up-scattered self-absorption energy is lowered by a factor 10 and the model flux at to 2 keV 
becomes compatible with BeppoSAX observations of GRB 990123. Parameterizing $b = 0.03\, b_{-1.5}$, 
we find that the GRB emission is produced at
\begin{equation}
  r_\gamma = 1.2\times 10^{15}\, x_1^{-2/3}\, \eta_2^{-1/2}\, b_{-1.5}^{-1/2}\; {\rm cm}
\label{rr}
\end{equation}
the lower limit on the outflow Lorentz factor is
\begin{equation}
 \Gamma \ge 630\;  x_1^{1/2}\, \eta_2^{-1/6}\, b_{-1.5}^{1/6}
\label{GGm}
\end{equation}
and the upper limit on the magnetic field is
\begin{equation}
 B \le 760\;  x_1^{3/2}\, \eta_2^{1/6}\, b_{-1.5}^{-1/6}\; {\rm G} \;.
\end{equation}

 The sub-pulse duration of equation (\ref{tDelta}) is now $t_\Delta \le 0.25\; x_1^{-5/3}\, 
\eta_2^{-1/6}\, b_{-1.5}^{-5/6}$ s, therefore the source intrinsic fluctuations are marginally 
resolved and the observed fluctuation amplitude of about 10\% requires that $\eta \simeq 100$, 
which is the canonical value chosen in the above equations.
Requiring that the sub-pulse duration $t_\Delta$ does not exceed the FWHM duration of a GRB pulse,
which is $t_\gamma \simeq 10$ s, leads to  
\begin{equation}
 b > 6 \times 10^{-4}\, x_1^{-2}\, \eta^{-1/5}
\label{bmin1}
\end{equation}
with $\eta \simg 1$ because, for $t_\Delta \simeq t_\gamma$, the GRB pulse should be a single 
emission episode.

 The comoving-frame electron density of the shocked fluid is
\begin{equation}
  n' = \frac{N_e}{4\pi\,r^2\,(\Delta'/\zeta)} \ge
        5.8\times 10^7\;\zeta\, x_1^{5/6}\, \eta_2^{1/3}\, b_{-1.5}^{2/3}\;\; {\rm cm^{-3}}
\label{nco}
\end{equation}
where $\zeta$ is the shock compression factor. Therefore, the magnetic field energy is a fraction
\begin{equation}
 \varepsilon_B = \frac{B^2/8\,\pi}{\Gamma'\, n'\, m_p c^2} \simeq
                 0.27\; x_1^{13/6}\, (\Gamma'-1)^{-1}\zeta^{-1} b_{-1.5}^{-1}
\end{equation}
of the energy density in the shocked fluid, where $\Gamma'$ is the Lorentz factor of the 
shock energizing the GRB-emitting fluid measured in the frame of the yet unshocked plasma.
If the GRB ejecta were not initially highly magnetized then a sub-equipartition magnetic 
field ($\varepsilon_B < 0.5$) requires that $b > 0.03\, x_1^{13/6}\, (\Gamma'-1)^{-1} \zeta^{-1}$. 
For a relativistic shock with $\Gamma'\sim$ few and $\zeta = 4\Gamma'$, this condition 
becomes 
\begin{equation}
 b \simg 10^{-4}\, x_1^{13/6}
\label{bmin2}
\end{equation}
which is close to that obtained by requiring that $t_\Delta \siml 10$ s (equation \ref{bmin1}). 

 Thus, we find that the magnetic field length-scale is a fraction $b = 10^{-3.5}-10^{-1.5}$ 
of the thickness of the shocked gas. For the comoving-frame density given in equation 
(\ref{nco}), the plasma skin-depth in the shocked gas is
\begin{equation}
 \lambda = c \left( \frac{\pi\, \gamma_p\, m_e} {e^2\, n'} \right)^{1/2} \ge 
           1.1 \times 10^3 \; x_1^{-2/3}\, \eta_2^{-1/6} \, b_{-1.5}^{-1/3} \; {\rm cm} 
\end{equation}
thus the magnetic field decay length-scale, $b(\Delta'/\zeta)$, is $5\times 10^5 - 10^7$ 
times larger than the plasma skin-depth. 

 Lastly, we note that radius at which the expansion of the ejecta is affected by the 
interaction with the circumburst medium, obtained by using the GRB ejecta Lorentz factor 
$\Gamma$ instead of the Lorentz factor of the shocked medium, $\tilde{\Gamma}$, in equation 
(\ref{adb}), is
\begin{equation}
  r_{dec} \siml 
   1.4\times 10^{15}\,E_{54.5}\,A_*^{-1}\,x_1^{-1}\,\eta_2^{1/3}b_{-1.5}^{-1/3}\;{\rm cm} 
\end{equation}
which is close to the radius $r_\gamma$ where the burst emission is produced (equation \ref{rr}). 
This shows that, if the burst mechanism were internal shocks in a variable wind, the dynamics 
of these internal shocks is affected by the deceleration of the outflow and, perhaps, a large 
number of collisions are between ejecta shells and the decelerating leading edge of the outflow.
That the deceleration radius is comparable with the prompt emission radius is consistent with 
the subtraction of the back-extrapolated forward-shock emission from the optical prompt flux 
done for Figures \ref{t30} and \ref{t30a}: in the $r_{dec} > r_\gamma$ case, the power-law decay
of the forward-shock emission would set in only after $r_{dec}$ and the 0.1--2 day optical decay 
could be extrapolated backwards only up to an epoch which is after the burst.

\section{Conclusions}

 The underlying assumption of this work, that the ROTSE optical counterpart of GRB 990123 arose 
from the same mechanism as the burst, is motivated by the similarity of its timing and decay rate 
to those of the fast-decay phase of Swift X-ray afterglows. As the latter can be identified with 
the large-angle emission produced during the burst, we attribute the ROTSE optical counterpart to 
the same mechanism. However, the optical emission associated with GRB 990123 must be a different 
spectral component than the burst because the optical flux lies well above the extrapolation of 
the burst spectrum. In this way, we arrived at the synchrotron self-Compton model for GRB 990123 
and its optical counterpart.

 The spectral slope of the optical counterpart of GRB 990123 was not measured. Future observations 
of early optical afterglows will provide a very simple test of the large-angle emission for GRB 
optical counterparts: their power-law decay index and spectral slope should satisfy equation 
(\ref{alpha}). The synchrotron self-Compton interpretation of the optical and $\gamma$-ray
emissions of GRB 990123 implies that the optical spectral slope must be equal to either the
low-energy or the high-energy burst spectral slope. The decay index of the optical emission of 
GRB 990123 and the slope of the burst continuum below its peak satisfy equation (\ref{alpha}),  
thus providing support to the large-angle interpretation proposed for the optical counterpart.

 In the framework of the synchrotron self-Compton model, the ROTSE optical and BATSE $\gamma$-ray
observations for GRB 990123 allow us to determine that the radius at which the burst emission was 
produced is comparable to the outflow deceleration radius, which in itself does not rule out any
of the possible origins (internal, reverse-external, or forward-external shocks) of the burst
emission, but points to that, if the burst arises from internal shocks, then most of these shocks 
must have occurred on the decelerating, leading front of the outflow, as proposed by Fenimore 
\& Ramirez-Ruiz (1999). Alternatively, that the burst emission was produced at the deceleration 
radius gives support to the electromagnetic model of Lyutikov \& Blandford (2003), which predicted 
such a burst location.

 The outflow parameters derived from the optical and $\gamma$-ray properties of GRB 990123 imply 
an up-scattered self-absorption frequency of 60 keV, which is inconsistent with BeppoSAX observations, 
showing an optically-thin burst spectrum above 2 keV. This difficulty can be overcome if the magnetic 
field does not occupy the entire shocked gas. The magnetic field decay length-scale is upper-bound 
by the condition that the burst spectrum is optically thin above 2 keV and lower-bound by that the 
shell shock-crossing time should not be longer than the duration of a GRB pulse and the magnetic 
field energy should not exceed equipartition. From these conditions, we find that the magnetic field 
must occupy $10^{-3}-10^{-2}$ of the shocked shell, which is equal to $10^6-10^7$ plasma skin-depths. 
It is rather puzzling that the magnetic field decay-length is so much larger than the natural scale 
for magnetic field generation, and yet does not occupy the entire shell of shocked gas. Whether such
a large magnetic field decay length-scale is possible remains an open question which cannot be
currently addressed by numerical models of two-stream instabilities (Medvedev \& Loeb 1999), due to 
the large computational effort required to follow the evolution of magnetic fields over such long 
scales. We note that other researchers have obtained similar constraints on the magnetic field decay
length-scale: from energetic arguments related to the outflow parameters obtained through afterglow 
modelling, Rossi \& Rees (2003) have set a low limit of $10^{-2}$ on the fraction of shell filled by 
magnetic field, while Pe'er \& Zhang (2006) have inferred a decay length-scale smaller by a factor 
10 than our value, from the condition that, in internal-shocks synchrotron-emission GRBs, electrons 
do not cool significantly during the burst. 

 There are two other bursts whose accompanying optical emission has been measured.
 The optical and $\gamma$-ray light-curves of GRB 041219A (Vestrand \etal 2005) are correlated
and the post-burst decay of the optical counterpart exhibits variability (Blake \etal 2005), 
both indicating that the counterpart is not the large-angle emission released during the burst.
 The optical and $\gamma$-ray emissions of GRB 050820A (Vestrand \etal 2006, Cenko \etal 2006) 
are not correlated, consistent with the large-angle emission scenario, but the $t^{-1}$ post-burst 
decay of the optical afterglow is too slow for that interpretation: the burst spectrum, $F_\nu 
\propto \nu^{-0.1\pm0.1}$, and equation (\ref{alpha}) imply a steeper, $t^{-2}$ decay for the 
large-angle emission. Therefore, GRB 990123 is so far the only case exhibiting a fast-decaying
optical counterpart, uncorrelated with the burst emission, that can be interpreted as arising
from the same mechanism as the burst.


\end{document}